\documentclass[12pt]{article}
\def\fun#1#2{\lower3.6pt\vbox{\baselineskip0pt\lineskip.9pt
\ialign{$\mathsurround=0pt#1\hfil ##\hfil$\crcr#2\crcr\sim\crcr}}}
\def\la{\mathrel{\mathpalette\fun <}}
\newcommand{\vep}{\mbox{\boldmath${\rm p}$}}
\newcommand{\lan}{\langle}
\newcommand{\ran}{\rangle}
\begin{document}
\begin{center}

{\Large \bf The $\mathbf{S}-\mathbf{D}$ mixing and di-electron
widths of higher charmonium $\mathbf{1^{--}}$ states}

\vspace{1cm}

\large{A.M.~Badalian}\footnote{ Institute of Theoretical and
Experimental Physics, Moscow, Russia.}$^*$, \large{
B.L.G.~Bakker}\footnote{Department of Physics and Astronomy, Vrije
Universiteit, Amsterdam, The Netherlands.},
\large{I.V.~Danilkin}\footnote{Moscow Engineering Physics
Institute, Moscow, Russia.\\$^*$ E-mail: badalian@itep.ru}
\end{center}
\vspace{1cm}

\begin{abstract}
\noindent The di-electron widths of $\psi(4040)$, $\psi(4160)$,
and $\psi(4415)$, and their ratios are shown to be in good
agreement with experiment, if in all cases the $S-D$ mixing with a
large mixing angle $\theta\approx 34^\circ$ is taken. Arguments are
presented why continuum states give small contributions to the
wave functions at the origin. We find that the $Y(4360)$
resonance, considered as a pure $3\,{}^3D_1$ state, would have
very small di-electron width, $\Gamma_{ee}(Y(4360))=0.060$ keV. On
the contrary, for large mixing between the $4\,{}^3S_1$ and
$3\,{}^3D_1$ states with the mixing angle $\theta=34.8^\circ$,
$\Gamma_{ee}(\psi(4415))=0.57$ keV coincides with the experimental
number, while a second physical resonance, probably $Y(4360)$, has
also a rather large $\Gamma_{ee} (Y(\sim 4400))=0.61$ keV. For the
higher resonance $Y(4660)$, considered as a pure $5\,{}^3S_1$
state, we predict the di-electron width
$\Gamma_{ee}(Y(4660))=0.70$ keV, but it becomes significantly
smaller, namely $0.31$ keV, if the mixing angle between the
$5\,{}^3S_1$ and $4\,{}^3D_1$ states $\theta=34^\circ$. The mass
and di-electron width of the $6\,{}^3S_1$ charmonium state are
calculated.

\end{abstract}

\section{Introduction}
\label{sect.1}

Knowledge of the di-electron widths of higher charmonium states is
important for many reasons. First of all, it can help to identify
the nature of the newly discovered resonances with $J^{PC}=1^{--}$
and distinguish between conventional $c\bar c$ mesons and, for
example, tetraquarks which have much smaller di-electron widths
\cite{1}. As shown in \cite{1}, di-electron widths of compact
four-quark systems, like $c\bar cq\bar q$, with $J^{PC}=1^{--}$
appear to be two orders smaller than those of conventional $c\bar
c$ mesons.

For higher charmonium states, which lie above the $D\bar D^*$ (or
$D^*\bar D^*$) threshold, their wave functions (w.f.) can be
strongly affected by the coupled-channel and threshold effects,
being in general very complicate functions. Via open channels the
w.f. of a resonance contains admixtures of other states with the
same quantum numbers and also a contribution from continuum
state(s). To define such a w.f. one needs to formulate the
relativistic many-channel Hamiltonian in QCD, even if in some
approximation \cite{2}. In the simplest case, the continuum part
consists of two open-charm mesons and to some extent this
continuum part can be considered as a particular case of a
four-quark system, $c\bar c q\bar q$. Since the contribution of
any four-quark state to the w.f. at the origin is much smaller
than that of a meson, it can be neglected. This effect occurs
because the probability to collect four (and even three) particles
at the origin is much smaller than for two particles. This fact
does not exclude that at larger distances the continuum part can
give an essential contribution to the w.f. and even dominates
asymptotically. Just such a coupling to the continuum provides a
shift down of the mass of the $P$-wave heavy-light mesons \cite{2}.
Therefore the w.f. at the origin of a higher vector resonance,
which we are interested in here, can be calculated taking into
account only the mixing between those vector states which masses,
defined in single-channel approximation, have close values. Study
of the charmonium spectrum shows that the $(n+1)\,{}^3S_1$ and
$n\,{}^3D_1$ states ($n\geq 3$) have small mass differences, $\leq
60$ MeV, which decrease for higher radial excitations. On the
other hand the mass differences between neighbouring $^3S_1$
states is of the order of several hundreds of MeV, so in first
approximation mixing between these states can be neglected. Such a
representation of the w.f. at the origin can be tested via
concrete predictions for the di-electron widths of different
vector states in heavy quarkonia.

In this picture the $S-D$ mixing between higher resonances can be
considered in the same way as it has been done for $\psi'(3686)$
and $\psi''(3770)$ \cite{3}, \cite{4}, where the mixing angle
$\theta =(12\pm 2)^\circ$ is extracted from the ratio of their
di-electron widths. Here we show that for higher vector states the
S-D mixing is significantly larger and the mixing angle
$\theta\sim 34^\circ$. Just for such an angle the di-electron
widths of $\psi(4040)$ and $\psi(4160)$ turn out to be almost
equal, as in experiment \cite{5}-\cite{9}.

There are also other arguments in favor of a large $S-D$ mixing.
It is known that the experimental di-electron widths of
$\psi(4040)$, $\psi(4160)$, and $\psi(4415)$ are significantly
smaller than the potential model predictions in single-channel
approximation \cite{10}-\cite {13}. Moreover, even with the use of
a many-channel w.f. at the origin (calculated in \cite{12} for the
Cornell coupled-channel model \cite{14}) the di-electron widths of
higher charmonium states appeared to be considerably larger than
in experiment. It is well-known that the QCD-motivated
gluon-exchange (GE) potentials (with the strong coupling,
possessing the asymptotic freedom property) have much smaller w.f. 
at the origin than those with $\alpha_s=const$ \cite{13};
nevertheless, even such potentials give di-electron widths of
$\psi(4040)$ and $\psi(4415)$ which are still $50-70\%$ larger
than the experimental numbers \cite{10}-\cite{12}. In the recent
paper \cite{15} the $Y(4660)$ resonance \cite{16}, considered as a
$5\,{}^3S_1$ state, has also large $\Gamma_{ee}(5\,{}^3S_1)=1.34$
keV, which is even significantly larger than
$\Gamma_{ee}(\psi(4415))=0.58(7)$ keV.

We assume here that the relatively small values of the di-electron
widths of the $n\,{}^3S_1$ states and the rather large widths of the
$n\,{}^3D_1$ states, initially considered as pure states, occur
mostly due to $S-D$ mixing. Here we do not use the same assumption
as made in \cite{11}, where to reach agreement with experimental
widths, ``total" screening of the GE interaction at large
distances has been supposed: such an assumption cannot explain why
$\Gamma_{ee}(\psi(4160))$ is large, and also has no deep
theoretical grounds.

Three experimental facts point to a possibly large mixing between
the
$(n+1)\,{}^3S_1$ and $n\,{}^3D_1$ vector charmonium states:

\begin{enumerate}

\item
The measured di-electron width of $\psi(4160)$, which is usually
considered as the $2\,{}^3D_1$ state, is large \cite{8}, \cite{9}:
\begin{equation}
\label{1}
\Gamma_{ee}(\psi(4160))=0.83\pm 0.06~ {\rm keV}.
\end{equation}
Namely, it is only $5-10\%$ smaller than the di-electron width of
$\psi(4040)$ and $\sim$14 times larger than the width calculated
here for a pure $2\,{}^3D_1$ state:
$\Gamma_{ee}(2\,{}^3D_1)=0.061$ keV (see Section 3). It is also
about three times larger than $\Gamma_{ee}(\psi''(3770))=
0.248(6)$ keV.

\item
On the contrary, the experimental width of $\psi(4040)$,
considered usually as the $3\,{}^3S_1$ state,
\begin{equation}
\label{2}
\Gamma_{ee}(\psi(4040))=0.86\pm 0.07~{\rm keV},
\end{equation}
appears to be almost two times smaller than for a pure
$3\,{}^3S_1$ state \cite{12}.

This situation can be resolved if the S-D mixing between these two
states is taken into account. For levels above the $D^*\bar D^*$
threshold such a mixing can occur owing to short-range tensor
forces (it gives a rather small effect) and the influence of open
channel(s). Since at present there is no dynamical calculation of
$S-D$ mixing, the influence of open channels can be taken into
account in a phenomenological way through the introduction of a
mixing angle, as for $\psi'(3686)$ and $\psi''(3770)$ \cite{3},
\cite{4}.

\item
The third fact refers to the di-electron width of $\psi(4415)$. If
this resonance is considered as the $4\,{}^3S_1$ state, then
potential models give di-electron widths in the range $1.1-1.5$
keV \cite{10}, \cite{12}, \cite{13}, which are almost two times
larger than in experiment \cite{8}, \cite{9}:
\begin{equation}
\label{3}
\Gamma_{ee}(\psi(4415))=0.58\pm 0.07~{\rm keV}.
\end{equation}
Such a decrease of the di-electron width could occur via mixing
with a still unidentified $3\,{}^3D_1$ state, which in
single-channel calculations has mass $M(3D)= 4.470(10)$ MeV, while
$M(4S)=4420(10)$ MeV, i.e., these two masses are rather close to
the masses of the physical resonances $\psi(4415)$ and $Y(4360)$.
One may expect that these $4\,{}^3S_1$ and $3\,{}^3D_1$ states
could be strongly coupled to the $S$-wave decay channels, like
$D_1(2420)D^*(2010)$, $D_0^*(2400) D^*(2010)$, and
$D_{s0}^*(2317)D_s^*(2112)$, and due to this coupling the
$4\,{}^3S_1$ and $3\,{}^3D_1$ levels are mixed and acquire hadronic
downward mass shifts, which are typically $\sim 40-60$ MeV
\cite{14}. As a result, one of the shifted physical states goes
over into the conventional $\psi(4415)$ charmonium, while the
other one can possibly be identified with the $Y(4360)$ resonance,
recently discovered by the Belle Collaboration \cite{16}. (In our
analysis here, the Belle resonance $Y(4360)$ with $\Gamma=48(15)$
MeV \cite{16} and the wide resonance $Y(4324)$ with $\Gamma=172$
MeV, observed by the BaBar collaboration \cite{17}, are considered
to be the same). Then the di-electron width of $\psi(4415)$ is
calculated here, taking into account large $S-D$ mixing, while for
the analysis it is inessential from which state, $4\,{}^3S_1$ or
$3\,{}^3D_1$, the resonance $\psi(4415)$ originates. We show that
for $\theta=34^\circ$ the di-electron widths of both physical
resonances have close values: $\Gamma_{ee}(Y(4360))\sim
\Gamma_{ee}(\psi(4415))=0.58$ keV.

\end{enumerate}

In our picture it is convenient to define the mixing angle between
higher vector states from the ratio of the di-electron widths, as
in \cite{3}, \cite{4}: in this case the QCD factor $\beta_V$,
occuring due to radiative corrections (see Sect.~\ref{sect.4}), is
cancelled in the ratio. From such an analysis a large mixing angle
is extracted and the absolute values of $\Gamma_{ee}(\psi(4040))$
and $\Gamma_{ee}(\psi(4160))$ are obtained in good agreement with
experiment if the same QCD factor $\beta_V=0.63$ is taken for all
higher states.

Notice that the mass of the $5\,{}^3S_1$ state,
$M(5\,{}^3S_1)=4640(10)$ MeV, has been predicted in \cite{12},
before the Belle resonance $Y(4660)$ was discovered \cite{16}.
This resonance and its radiative transitions were studied in
detail in \cite{15} giving $\Gamma_{ee}(Y(4660))=1.34$~keV. In our
calculations $\Gamma_{ee}(Y(4660))$ strongly depends on a possible
admixture of the $4\,{}^3D_1$ state and is considerably smaller
than in \cite{15}: $\Gamma(Y(4660))=0.70$ keV, if this resonance
is a pure $5\,{}^3S_1$ state, and about two times smaller,
$\Gamma_{ee}(Y(4660))=0.31$ keV, if the mixing angle between
$5\,{}^3S_1$ and the unobserved $4\,{}^3D_1$ state (with mass
$\sim 4700$ MeV) is $34^\circ$, the same as for $\psi(4415)$.

\section{The masses of the ${{J^{PC}=1^{--}}}$ charmonium states}
\label{sect.2}

The hyperfine (HF) and fine-structure splittings of higher radial
excitations are small ($\leq 20$ MeV) \cite{11}, \cite{18},
therefore their masses practically coincide with the centroid
masses, $M_{\rm cog}(nL)$, which we need to determine with good
accuracy. To calculate them we use here the relativistic string
(RS) Hamiltonian with universal (for all mesons) interaction
\cite{19}, \cite{20}. For charmonium one contribution to the mass
formula, namely the small string correction ($\leq5$ MeV ) for the
states with $L\not=0$ can be neglected, while the self-energy
correction, $\sim -20$ MeV, is taken into account here.

In heavy quarkonia the mass $M_{\rm cog}(nL)$ is just given by the
eigenvalue (e.v.) of the spinless Salpeter equation (SSE)
\cite{20}:
\begin{equation}
\label{4}
\left\{ 2\sqrt{\vep^2 +m^2_c}+ V_B(r)\right\} \psi_{nL} (r) =
M_{\rm cog} (nL) \psi _{nL} (r).
\end{equation}
Here we also use the RS Hamiltonian written in the Einbein
approximation (EA) \cite{21}. In this case the spin-averaged mass
can be presented as:
\begin{equation}
\label{5}
M_{\rm cog}(nL)=\omega_{nL}+\frac{m_c^2}{\omega_{nL}}+
E_{nL}(\omega_c)+\Delta_{SE},
\end{equation}
where the e.v. $E_{nL}$ are the solutions of the so-called einbein
equation \cite{21}, \cite{22}:
\begin{equation}
\label{6}
\left[\frac{\vep^2}{\omega_{nL}}+V_0(r)
\right]\varphi_{nL}(r)=E_{nL}\varphi_{nL},
\end{equation}
which together with the mass $\omega_{nL}$ should be defined in a
selfconsistent way:
\begin{equation}
\label{7}
\omega^2_{nL}=m^2_c-\frac{\partial E_{nL}}{\partial \omega_{nL}}.
\end{equation}

For the $n\,{}^3D_1$ state the small HF contribution to the mass
will be neglected here, i.e., its mass $M(n\,{}^3D_1)=M_{\rm
cog}(nD)$, while for the $n\,{}^3S_1$ states we still keep the
small HF correction: $M(n\,{}^3S_1)=M_{\rm cog} (nS) +
\frac{1}{4}\delta_{\rm HF}(nS)$ with $\delta_{\rm HF} (nS) =
M(n\,{}^3S_1) - M((n^1S_0)$. The values of $\delta_{\rm
HF}(nS)=48(48)$, 16(20), 12(16), 6(10) MeV $(n=2,3,4,5)$,
calculated in \cite{11} and \cite{18} (in parentheses), are used
here.

Our calculations are performed with the universal potential
$V_B(r)$ from \cite{20}, \cite{22}:
\begin{equation}
\label{8}
V_B(r) =\sigma(r) \cdot r-\frac43 \frac{\alpha_B(r)}{r},
\end{equation}
where the vector coupling $\alpha_B(r)$ is taken in two-loop
approximation: it has the asymptotic freedom behavior at small
$r$, freezes (saturates) at large $r$, and depends on the number
of flavors $n_f$. For charmonium we use $n_f=4$ and the QCD
constant $\Lambda_{\overline{\rm MS}}^{(4)}= 254$ MeV, which gives
the vector QCD constant $\Lambda_V (n_f=4)
=1.4238\cdot\Lambda_{\overline{\rm MS}}(n_f=4)=360$ MeV \cite{22}.
The freezing (critical) value of $\alpha_B(r)$ is expressed
through $\Lambda_V$ and the so-called background mass $M_B=1.0$
GeV:
\begin{equation}
\label{9}
\alpha_B(r\to \infty) =\alpha_B(q=0) =\frac{4\pi}{\beta_0 t_0}
\left(1 -\frac{\beta_1}{\beta^2_0} \frac{\ln t_0}{t_0}\right) = 0.546,
\end{equation}
where $t_0=\ln \left(\frac{M_B}{\Lambda_V}\right)2$, $\beta_0
=11-\frac23 n_f$, and $\beta_1=102 -\frac{38}{3} n_f$.

For low-lying states which have relatively small sizes (with
r.m.s. radius $R(nL)\leq0.8$ fm), a linear confining potential
with constant string tension, $\sigma=\sigma_0\cong 0.18$ GeV$^2$,
can be used \cite{23}. However, for higher states, which lie above
open thresholds and have large radii $R(nL)\geq 1.0$ fm, it is
important to take into account the creation of virtual light-quark
pairs $(q\bar q)$, even in single-channel approximation. Due to
virtual loops the surface inside the Wilson loop decreases, making
the string tension dependent on the $Q \bar Q$ separation $r$
\cite{24}:
\begin{equation}
\label{10}
\sigma(r) =\sigma_0 (1-\gamma f(r)).
\end{equation}
Such a flattening of the confining potential is common to all
mesons of large sizes, and therefore for charmonium the form and
parameters of such amodified string tension can be taken from the
analysis of the radial Regge trajectories for light mesons
\cite{24}:
\begin{equation}
\label{11}
\gamma=0.40;~~ f(r\to 0) =0,\quad f(r\to\infty) =1.0.
\end{equation}
As shown in \cite{24}, due to flattening the masses of all higher
levels are shifted down and these mass shifts increase with $n$.
For example, the shift of the $5\,{}^3S_1$ state reaches
$\sim$~100 MeV \cite{23}. In Tables~\ref{tab.1} and \ref{tab.2} we
give the masses of pure $n\,{}^3S_1$ and $n\,{}^3D_1$ states for
the potential (\ref{8}), which are calculated using the SSE and
the EA (\ref{6}), and also the masses calculated in \cite{18},
where in the Cornell potential a constant coupling is used, equal to our
freezing value (\ref{9}).
\begin{table}
\caption{The charmonium masses $M(n\,{}^3S_1)$ (in MeV) for the
potential (\ref{8}).\label{tab.1}}
\begin{center}
\begin{tabular}{|l|l|l|l|l|}
\hline
state & SSE$^{a)}$ & EA$^{a)}$& BGS[18] & exp.\cite{9}\\
& $m_c=1.425$ GeV & $m_c=1.410$ GeV & $m_c=1.4794$ GeV& \\
\hline
 $1S$ & 3105 & 3095 & 3090 & 3097\\
\hline
 $2S$ & 3678 & 3682 & 3672 & 3686\\
\hline
 $3S$ & 4078 & 4096 & 4072 & 4039(1)\\
\hline
 $4S$ & 4398 & 4426 & 4406 & 4421(4)\\
& & & & 4361(18)$^{b)}$\\
\hline
 $5S$ & 4642 & 4672 & & 4664(16)$^{b)}$\\
\hline
 $6S$ & 4804 & 4828 & &\\
\hline
\end{tabular}
\end{center}

${}^{a}$ The self-energy corrections to the masses,
$\Delta_{SE}(nS)\approx -20$ MeV, are taken into account both in
the relativistic case (SSE) and in the
einbein approximation.\\
${}^{b}$ Belle data \cite{16}

\end{table}
From Table~\ref{tab.1} one can see that in our calculations the
mass of the $5\,{}^3S_1$ state agrees with that of the $Y(4660)$
resonance \cite{16}. For the $6\,{}^3S_1$ level the predicted mass
is $M(6\,{}^3S_1)=4815(15)$ MeV.

As seen in Tables~\ref{tab.1} and \ref{tab.2}, the masses
$M((n+1)S)$ and $M(nD)$ are close to each other, even in
single-channel approximation. The difference between them
decreases for larger radial excitations, so that for $n=5$ it is
only $\sim 30$ MeV.
\begin{table}
\caption{The charmonium masses $M(n\,{}^3D_1)$ (in MeV) for the
potential (\ref{8}).\label{tab.2}}
\begin{center}
\begin{tabular}{|l|l|l|l|l|}
\hline
state & SSE$^{a)}$ & EA$^{a)}$& BGS[18] & Exp. \cite{9}\\
&$m_c=1.425$ GeV & $m_c=1.410$ GeV &$m_c=1.4794$ GeV & \\
\hline
 $1D$ & 3800 & 3779 & 3806 &3770(3)\\
\hline
 $2D$ & 4156 & 4165 & 4167 &4159(3)\\
\hline
 $3D$ & 4464 & 4477 & &4421 \\
& & & &4361$^{b)}, 4324^{c)}$\\
\hline
 $4D$ & 4690 & 4707 & &\\
\hline
 $5D$ & 4840 & 4855 & &\\
\hline
\end{tabular}
\end{center}
$^{a}$ See the footnote $^{a}$ to Table~\ref{tab.1}\\
$^{b}$ See the footnote $^{b}$ to Table~\ref{tab.1}\\
$^{c}$ BaBar data \cite{17}
\end{table}
It is worthwhile to notice that besides the ``correlated" mass
shifts of higher levels---due to virtual pair creation---some
levels, which lie near thresholds, can have additional downward
shifts due to strong coupling to a continuum channel. We denote
these mass shifts as decay-channel (DC) shifts; they can be
calculated only within a multi-channel approach. The masses of the
$2S$, $4S$, and $2D$ levels calculated here (see
Tables~\ref{tab.1} and \ref{tab.2}) agree with experiment within
$20$ MeV, i.e., they have essentially no DC shifts. However, the
mass of the $3\,{}^3S_1$ level is $\sim 40$ MeV larger than the
experimental one, because this level can be affected by the
$D^*\bar D^*$ channel and we estimate its DC shift as $\sim 40$
MeV. Therefore, the masses predicted in our paper, have different
accuracies, which is better than 20 MeV for the levels without DC
shifts, and than 40 MeV for the levels strongly coupled to nearby
continuum decay channels.

Thus from our analysis we conclude that the mass difference,
\begin{equation}
\label{12}
\Delta_n M= M(n\,{}^3D_1) -M((n+1)\,{}^3S_1),
\end{equation}
decreases from the value $\Delta_2M(\exp) =120$ MeV for $n=2$
to $\sim 30$ MeV for $n=4$. Therefore higher levels are almost
degenerate and the $S-D$ mixing for them, as well as the mixing
angle, become larger than for $\psi'$ and $\psi''$. Also in
single-channel approximation $M_{\rm cog}((n+1)S)$ is always
smaller than $M_{\rm cog}(nD)$.

\section{Vector Decay Constants}
\label{sect.3}

The decay constants of vector (V) mesons are calculated here using
the analytical expressions derived in \cite{25} in the framework
of the Field Correlator Method, where $f_V(nS)$ and $f_V(nD)$ are
given by
\begin{equation}
\label{13}
 f^2_V(nS)=12\frac{\left|\psi_{nS}(0)\right|^2}{M_V(nS)}\,
 \xi_V= \frac{3}{\pi}\frac{\left|R_{nS}(0)\right|^2}{M_V(nS)}\, \xi_V,
\end{equation}
where the relativistic factor $\xi_V$ is defined as
\begin{equation}
\label{14}
 \xi_V(nL)=\frac{m^2+\omega_{nL}^2+
 \frac{1}{3}\langle\vep^2\rangle}{2\omega_{nL}^2}.
\end{equation}
The same expression (\ref{13}) can be applied to the decay
constants $f_V(n\,{}^3D_1)$, if in (\ref{13}) the w.f. $R_{nD}(0)$
is defined as in (\ref{15}) below.

All matrix elements (m.e.) which are needed to calculate the decay
constants for the $nS$ and $nD$ states, are given in
Tables~\ref{tab.3} and \ref{tab.4}. An interesting fact is that
for the $(n+1)S$ and $nD$ states the m.e. like $\omega_{nL}$,
$\langle\vep^2\rangle_{nL}$, and $\xi_V(nL)$ coincide with an
accuracy better than $1\%$ and therefore the difference between
$f_V((n+1)S)$ and $f_V(nD)$ comes only from their w.f. at the
origin and the small differences in their masses $M_V(nL)$.

\begin{table}[ht]
\caption{The matrix elements $\omega_{nS}$
(GeV),~$\langle\vep^2\rangle$ ((GeV$/c)^2$), and the w.f. at the
origin $R_{nS}(0)$ (GeV$^{3/2}$) (no mixing) for the potential
(\ref{8})\label{tab.3}} \center
\begin{tabular}{|c|c|c|c|c|}
\hline State & $\omega_{nS}$ (GeV) & $R_{nS}(0)$ &
$\langle\vep^{\,2}\rangle$ &
$\xi_{nS}$\\
\hline\hline
 $1S$ & 1.59 & 0.905 & 0.541 & 0.929 \\
 $2S$ & 1.65 & 0.767 & 0.722 & 0.910\\
 $3S$ & 1.69 & 0.714 & 0.882 & 0.899 \\
 $4S$ & 1.71 & 0.655 & 0.947 & 0.894\\
 $5S$ & 1.66 & 0.531 & 0.775 & 0.908\\
 $6S$ & 1.63 & 0.445 & 0.665 & 0.916\\
\hline
\end{tabular}
\end{table}

\begin{table}[ht]
\caption{The matrix elements $\omega_{nD}$ (GeV),
$\langle\vep^2 \rangle$ ((GeV$/c)^2$), the w.f. at the origin
$R_{nD}(0)$ (GeV$^{3/2}$), and the second derivative $R_{nD}''(0)$
(GeV${}^{7/2}$) for the potential (\ref{8}).\label{tab.4}} \center
\begin{tabular}{|c|c|c|c|c|c|}
\hline
State & $\omega_{nD}$ (GeV) & $R_{nD}''(0)$ & $R_{nD}(0)$
& $\langle\vep^{\,2}\rangle$ & $\xi_{nD}$ \\
\hline\hline
 $1D$ & 1.65 & 0.145 & 0.095 &0.721 & 0.909 \\
 $2D$ & 1.69 & 0.213 & 0.132 & 0.881 & 0.899 \\
 $3D$ & 1.71 & 0.248 & 0.150 & 0.939 & 0.893 \\
 $4D$ & 1.65 & 0.221 & 0.144 & 0.745 & 0.911\\
 $5D$ & 1.64 & 0.206 & 0.135 & 0.682 & 0.912\\
\hline
\end{tabular}
\end{table}

The w.f. at the origin $R_{nD}(0)$ is defined here as in
\cite{26}, being expressed via the second derivative
$R_{nD}^{''}(0)$:
\begin{equation}
\label{15}
R_{nD}(0)=\frac{5}{2\sqrt{2}\omega_{nD}2}R_{nD}^{ \prime \prime}(0).
\end{equation}
It is interesting that $R_{nS}(0)$ and $R_{nD}(0)$ have different
behavior for growing $n$: while $R_{nS}(0)$ decreases for higher
radial excitations, the second derivative
$R_{nD}^{\prime\prime}(0)$ and $R_{nD}(0)$ grow with increasing
$n$, if a linear confining potential is used. For the flattening
potential used here, $R_{nS}(0)$ decreases even faster, while
$R_{nD}(0)$ increases for the $2S$ and $3S$ states and then
practically saturates for higher levels. This growth of
$R_{nD}(0)$ is possibly one of the reasons why higher radial
excitations have large $S-D$ mixing.

In Table~\ref{tab.5} the decay constants of the charmonium states
with $J^{PC}=1^{--}$ are given in two cases: without and with $S-D$
mixing.
\begin{table}[ht]
\caption{The decay constants $f_V(n\,{}^3S_1)$ (in MeV) without
and with $S-D$ mixing.\label{tab.5}}
\center
\begin{tabular}{|c|c|c|c|}
\hline
\multicolumn{4}{|c|}{$\theta=0$}\\
\hline
& $n=1$ & $n=2$ & $n=3$ \\
\hline
$f_V((n+1)\,{}^3S_1)$ & 373 & 329 & 288 \\
$f_V(n\,{}^3D_1)$ & 45 & 60 & 66 \\
\hline
\multicolumn{4}{|c|}{$\theta \neq 0$}\\
\hline
$\theta$ & 11$^\circ$ & 34.8$^\circ$ & 34$^\circ$ \\
\hline
$f_V(\psi_S)$ & 357 & 236 & 202 \\
$f_V(\psi_D)$ & 115 & 234 & 217 \\
\hline
\end{tabular}
\end{table}
The mixing angle between the $2\,{}^3S_1$ and $1\,{}^3D_1$ levels
has already been calculated in \cite{3} and \cite{4}, and also in
\cite{23}, where the mixing angle $\theta=11^\circ$ has been
extracted from the ratio of experimental di-electron widths,
$\Gamma_{ee}(\psi'(3686))$ and $\Gamma_{ee}(\psi^{''}(3770))$.
Other mixing angles are calculated below. The matrix elements and
other numbers in Tables~\ref{tab.3} and \ref{tab.4} are calculated
here with the use of the EA equation (\ref{6}), which provides
regular behavior of the w.f. at the origin, in contrast to the SSE
for which the $S-$wave w.f. diverges at the origin and needs to be
regularized.

As seen in Table~\ref{tab.5} the decay constants $f_V(nD)$ are
very small, $\sim 50$ MeV ($\theta=0$) for pure $D$-wave states.
However, if $S-D$ mixing is large ($\theta\sim 34^\circ$) the
decay constants $f_V(\theta)$ ($n\geq 2$) of physical, ``mixed"
states appear to be practically equal for $\psi(4040)$ and
$\psi(4160)$, and also for $\psi(4415)$ and the still unidentified
second charmonium state, which originates from the $2D$ level and
it is denoted below as $\tilde \psi(4415)$ (although its mass is
$\sim 4470(10)$ MeV in single-channel approximation). For large
$S-D$ mixing all decay constants lie in the range $220\pm 20$ MeV
($n=2,3$). Precisely this fact provides close values of the
di-electron widths of the $(n+1)S$ and $nD$ states.

\section{Di-electron widths}
\label{sect.4}

In \cite{23} the di-electron widths: $\Gamma_{ee}(J/\psi)$,
$\Gamma_{ee}(\psi (3686))$, and $\Gamma_{ee}(\psi(3770))$ have
been calculated with high precision, $\leq 5\%$, using the
theoretical formula where the di-electron width is expressed via
the decay constant (\ref{13}), containing the relativistic
correction $\xi_V$, and includes QCD radiative corrections (this
expression is the relativistic generalization of the van
Royen-Weisskopf formula \cite{27} in the framework of Field
Correlator Method). The QCD correction, known in one-loop
approximation, enters as the multiplicative factor denoted here as
$\beta_V=1-\frac{16}{3\pi}\alpha_s(M_V)$. Then
\begin{equation}
\label{16}
 \Gamma_{ee}(n\,{}^3S_1)=\frac{4 \pi e_c^2\alpha^2}{3
 M_{nS}}f_{nS}^2\beta_V=\frac{4e^2_c\alpha2}{M^2_{nS}}|R_{nS}(0)|^2\xi_{nS}
 \beta_V,
\end{equation}
\begin{equation}
\label{17}
 \Gamma_{ee}(n\,{}^3D_1)=\frac{4 \pi e_c^2\alpha^2}{3
 M_{nD}}f_{nD}^2\beta_V=\frac{4e^2_c\alpha^2}{M^2_{nD}}|R_{nD}(0)|^2\xi_{nD}
\beta_V.
\end{equation}
The w.f. at the origin $R_{nD}(0)$ in (\ref{17}) has been defined
in (\ref{15}) and the average kinetic energy $\omega_{nL}=\lan
\sqrt{\vep^2+m^2_c}\ran_{nL}$, calculated from equation (\ref{7}),
plays the role of a constituent quark mass being different for
different $nL$ states. The $\omega_{nL}$ have the following
characteristic feature: For a linear confining potential with
$\sigma =\sigma_0=const.$ it grows for higher radial excitations,
while for the flattening potential first it grows for $n=2$, 3,
and 4 and then saturates around the value $\omega_{nL}\sim 1.65$
GeV for $n\geq 5$ (the values $\omega_{nS}$, $\omega_{nD}$ are
given in Tables \ref{tab.3} and \ref{tab.4}).

The expressions (\ref{16}) and (\ref{17}) contain the $c$-quark
charge $e_c=2/3$, $\alpha=1/137$, and the mass $M_{nS}(M_{nD})$ of
the $n\,{}^3S_1(n\,{}^3D_1)$ vector mesons. The w.f. at the origin
$R_{nS}(0)$, $R_{nD}(0)$, and $R_{nD}''(0)$ are given in Tables
\ref{tab.3} and \ref{tab.4}.

As we discussed in the Introduction and Section 2, one may expect
that the physical $\psi$-mesons represent a mixing of the $(n+1)S$
and $nD$ states with close mass values, and our goal here is to
determine the mixing angle between higher radial excitations. To
this end we introduce the w.f. of the physical $\psi$-mesons,
denoted here by $\varphi_{nS}(0)$ and $\varphi_{nD}(0)$, where the
symbols $nS$ and $nD$ simply remind about the origin of those
states:
\begin{eqnarray}
\label{18}
\varphi_{nS}(0) & = & \cos\theta_nR_{nS}(0) - \sin\theta_nR_{(n-1)D}(0),
\nonumber \\
\varphi_{nD}(0) & = & \cos\theta_nR_{(n+1)S}(0) - \sin \theta_nR_{nD}(0).
\end{eqnarray}
The di-electron widths of the $\psi$-mesons are expressed via the
physical w.f. at the origin (\ref{18}) in the same way as in
(\ref{16}) and (\ref{17}). For a given mixing angle $\theta$ the
w.f. at the origin (\ref{18}) are easily calculated through the
w.f. $R_{nS}(0)$ and $R_{nD}(0)$ for pure $S$- and $D$-wave states
(they are given in Tables \ref{tab.3} and \ref{tab.4}).
\begin{table}[ht]
\caption{The wave functions at the origin $\varphi_{(n+1)S} (0)$
and $\varphi_{nD} (0)$ in GeV${}^{3/2}$ of the physical states for
$n=1,2,3,4^a$. \label{tab.6}}
\begin{center}
\begin{tabular}{|l|c|c|c|c|}
\hline
$n$ & $1$ & $2$ & $3$ & $4$\\
\hline
$\theta$ & $11^\circ$ & $34.8^\circ$ & $34^\circ$ & $34^\circ$\\
\hline
$\varphi_{(n+1)S} (0)$ & 0.735 & 0.511 & 0.459 & 0.360 \\
$\varphi_{nD} (0)$ & 0.240 & 0.516 & 0.491 & 0.416 \\
\hline
\end{tabular}
\end{center}
${}^a$ The uncertainty in the mass value used gives rise to a
theoretical error less than 1\%.
\end{table}

\section{$3\,{}^3S_1-2\,{}^3D_1$ mixing}
\label{sect.5}

To determine the mixing angle between the $3\,{}^3S_1$ and
$2\,{}^3D_1$ states we use here the ratio of the di-electron
widths of the $\psi(4040)$ and $\psi(4160)$ mesons as in \cite{4}
and \cite{22}. This ratio does not depend on the QCD factor
$\beta_V$ and the experimental di-electron widths are given in
(\ref{1}) and (\ref{2}) with their ratio close to unity. Such a
large ratio turns out to be possible only if the mixing angle
between $3\,{}^3S_1$ and $2\,{}^3D_1$ states is large.

Taking in the ratio the w.f. at the origin (\ref{18}), which are
expressed via the numbers $R_{nS}(0)$ and $R_{nD}(0)$ from Tables
\ref{tab.3} and \ref{tab.4}, one can extract the mixing angle
$\theta_2=34.8^\circ$ and determine the physical w.f. at the
origin, as well as the decay constants of $\psi(4040)$ and
$\psi(4160)$. Then for both charmonium states the di-electron
widths appear to be in precise agreement with experiment (see
(\ref{21})), if the QCD factor $\beta_V=0.63$ is taken. This value
of $\beta_V$ is smaller (i.e. the radiative corrections are
larger) than for $J/\psi$, $\psi(3686)$, and $\psi(3770)$, where
in all cases the larger value $\beta_V=0.72$ gives precise
agreement with experiment \cite{23}.

The mixing angle between the $(n+1)\,{}^3S_1$ and $n\,{}^3D_1$
states, denoted here as $\theta_n$, can be calculated if at least
one of the di-electron widths is known from experiment. For the
$3S$ and $2D$ states both di-electron widths are known and
$\theta_2$ is easily determined. It is important to notice that
for a pure $2\,{}^3D_1$ state the di-electron width is very small:
$\Gamma_{ee}(2\,{}^3D_1)$=0.059 keV, i.e., $\sim 14$ times smaller
than the experimental number (\ref{1}), and one can expect large
mixing between the $3\,{}^3S_1$ and $2\,{}^3D_1$ states. Such a
large mixing can occur via the nearby open $D^*\bar D^*$ channel
and partly through short-ranged tensor forces which, however, do
not provide a large mixing angle, $\theta({\rm tensor}) \la
7^\circ$. On the contrary, from the ratio:
\begin{equation}
\label{19}
\eta=\frac{\Gamma_{ee}(\psi(4040))}{\Gamma_{ee}(\psi(4160))}=1.04\pm0.17,
\end{equation}
one obtains two solutions with a large magnitude of $\theta_2$: a
positive and a negative one:
\begin{equation}
\label{20}
\theta_2=34.8^\circ \quad {\rm or} \quad\theta_2=-55.7^\circ .
\end{equation}
For these angles and using (\ref{18}) the physical w.f.
$\varphi(\psi(4040), r=0)=0.511$ GeV$^{3/2}$ and
$\varphi(\psi(4160), r=0)= 0.516$ GeV$^{3/2},$ appear to be almost
equal. Then from (\ref{16}) and (\ref{17}) with $\beta_V=0.63$ one
calculates the following di-electron widths:
\begin{equation}
\label{21}
\Gamma_{ee}(\psi(4040))=0.87~{\rm keV},\quad
\Gamma_{ee}(\psi(4160))=0.83~{\rm keV},
\end{equation}
which just coincide with the central values of the experimental
values, (\ref{1}) and (\ref{2}). The QCD factor $\beta=0.63$
extracted simultaneously, corresponds to the strong coupling
$\alpha_s(M_V)=0.217$. Later this value of $\beta_V=0.63$ is used
to determine $\theta_n$ for higher excitations ($n=3,4$). Notice
that the same mixing angle $\theta_2=35^\circ$ (or
$\theta_2=-55^\circ$) has been obtained in the analysis of
$\psi(4040)$ and $\psi(4160)$ \cite{28}.

\section{Large mixing between $4\,{}^3S_1$ and $3\,{}^3D_1$ states}
\label{sect.6}

In constituent quark models (in single-channel approximation) two
vector states, $4\,{}^3S_1$ and $3\,{}^3D_1$, are expected in the
mass region around 4.4 GeV (see Tables \ref{tab.1} and
\ref{tab.2}). Our calculations give the masses $M(4\,{}^3S_1)\sim
4.42$ GeV and $M(3\,{}^3D_1)\sim 4.47(1)$ GeV with their mass
difference $\sim 50$ MeV. However, one cannot exclude that due to
strong coupling to the $D^*D_1(2420)$ and $D^*D^*_2(2460)$
channels the $4\,{}^3S_1$ and $3\,{}^3D_1$ states are mixed and
the mass of one or probably both states is shifted down. Then one
of these mixed (physical) states can be identified with
$\psi(4415)$ and the other one with the newly discovered resonance
$Y(4360)$ \cite{16}. (Note that in charmonium the values of the
DC shifts are typically $\sim 40$ MeV \cite{14}).

From experiment only the di-electron width $\Gamma_{ee}(\psi(4415))$
is presently known, $\Gamma_{ee}(\psi(4415))= 0.58\pm 0.07$ keV,
while for the $Y(4360)$ resonance two possible numbers have been
measured for the product \cite{16},
\begin{equation}
\label{22}
B(Y(4360)\to \psi(2S) \pi^+\pi^-) \times
\Gamma_{ee}(Y(4360))=
\begin{array}{ll} a)&10.4\pm3.2~{\rm eV} \\b)&11.8\pm 3.2~{\rm eV}\end{array}
\end{equation}
Still, this restricted information allows one to draw an important
conclusion. First, for pure $4S$ and $3D$ states ($\theta_3=0$)
with the w.f. at the origin $R_{4S}(0)=0.65$ GeV$^{3/2}$ and
$R_{3D}(0) =0.150$ GeV$^{3/2}$ (from Tables \ref{tab.3} and
\ref{tab.4}), their di-electron widths are the following:
\begin{equation}
\label{22a}
\Gamma_{ee}(4\,{}^3S_1)=1.19~{\rm keV},\quad
\Gamma_{ee}(3\,{}^3D_1)=0.06~{\rm keV},\quad (\theta=0).
\end{equation}
i.e., $\Gamma_{ee}(4S)$ is two times larger than the experimental
number (\ref{3}) while $\Gamma_{ee}(3D)$ is small. To reach
agreement with experiment for $\psi(4415)$ we need to take a large
mixing angle, namely $\theta_3=34^\circ$, as for the $3S-2D$
mixing, for which
\begin{equation}
\label{23}
\Gamma_{ee}(\psi(4415))|_{\rm theory} =0.57~{\rm keV}
\end{equation}
is completely in agreement with the central experimental value
(\ref{3}).

Then for the same angle the di-electron width of the second
physical state, which can be denoted as $\tilde\psi(4470)$ (in
many-channel approximation its mass may be smaller), appears to be
ten times larger than for a pure $3\,{}^3D_1$ state:
\begin{equation}
\label{23a}
\Gamma_{ee}(\tilde\psi(4470))=0.63~{\rm keV}.
\end{equation}
Moreover this width is even slightly larger than that of
$\psi(4415)$ (here we take $\beta_V=0.63$ as for the $3S$ and $2D$
states). Since in this case the di-electron widths coincide within
$10\%$ accuracy, in the framework of the single-channel
approximation it is difficult to decide which of these states
should be identified with $\psi(4415)$ or with $Y(4360)$. From the
experimental value (\ref{22}) and the di-electron width (\ref{25})
one obtains an estimate of the branching $B(Y(4360)\to
\psi(2S)\pi^+\pi^-)$,
\begin{equation}
\label{21a}
B(Y(4360)\to \psi(2S)\pi^+\pi^-)\approx(1.6\pm 0.6)\%,
\end{equation}
which is rather large. Thus for large mixing angle $(\theta_3\cong
34^\circ)$ one cannot a priori conclude which resonance,
$\psi(4415)$ or $Y(4360)$, originates from the $4\,{}^3S_1$. A
decisive test for their identification could come from the study
of their radiative transitions to the $\chi_{cJ}$ states, because
radiative transitions are very sensitive to the mixing angle, as
has been shown for $\psi'(3686)$ and $\psi^{\prime\prime}(3770)$
in \cite{4}, \cite{29}.

\section{$Y(4660), Y(4815)$}
\label{sect.7}

The higher level $Y(4660)$ with $M=4660\pm 16$ MeV, recently
discovered in \cite{16}, has a surprisingly small width,
$\Gamma=48\pm 18$ MeV. This state lies close to the $S$-wave
threshold $D^*_sD_{s1}(2535)$ (with the threshold mass $M_{\rm
th}=4647$ MeV) and to the $P$-wave threshold $D(2\,{}^3S_1)\bar
D^*$ with $M_{\rm th}=4647$ MeV (our calculations give the mass
$M(D(2\,{}^3S_1))\approx 2640$ MeV). Our predictions for the
masses of the $5\,{}^3S_1$ and $4\,{}^3D_1$ states in
single-channel approximation (see Tables \ref{tab.1} and
\ref{tab.2}) are
\begin{equation}
\label{24}
M(5\,{}^3S_1)=4655(15)~{\rm MeV},\quad M(4\,{}^3D_1)=4700(10)~{\rm MeV},
\end{equation}
where the theoretical uncertainty is taken into account. These
masses differ only $\sim 50$ MeV and large mixing between the two
states can be expected. Unfortunately, at present their
di-electron widths remain unknown, and here we calculate them,
assuming that the QCD factor $\beta_V=0.63$ and
$\theta_4=34^\circ$ as it takes place for $4S$ and $3D$ states,
and also first consider pure $5S$ and $4D$ states $(\theta_4=0)$,
for which
\begin{equation}
\label{25}
\Gamma_{ee} (5\,{}^3S_1) =0.73\quad {\rm keV},\quad\Gamma_{ee} (4\,{}^3D_1)
=0.055~{\rm keV},~(\theta_4=0)
\end{equation}
i.e., $\Gamma_{ee}(5S)$ is rather large, being even larger than
$\Gamma_{ee}(4S)$ (\ref{3}), On the contrary $\Gamma_{ee}(4D)$ is
very small.

For large $S-D$ mixing with $\theta_4=34^\circ$ (like $\theta_2$
and $\theta_3$) we obtain
\begin{equation}
\label{26}
\Gamma_{ee}(\tilde\psi(4660))=0.32~{\rm keV},\quad (\theta_4=34^\circ)
\end{equation}
which is two times smaller than (\ref{25}). For the second state,
denoted as $\tilde\psi(4690)$ we find,
\begin{equation}
\label{27}
\Gamma_{ee}(\tilde\psi(4690))=0.45~{\rm keV},\quad (\theta_4=34^\circ)
\end{equation}
the width appears to be eight times larger than for the pure $4D$
state in (\ref{25}) and even larger than
$\Gamma_{ee}(\psi(4660))$. Notice that equal widths are obtained
for a bit smaller angle, $\theta_4 =30^\circ$:
\begin{equation}
\label{28}
\Gamma_{ee}(\tilde\psi(4660))=\Gamma_{ee}(\tilde\psi (4690))=0.39~{\rm
keV} .
\end{equation}
We predict also the $6\,{}^3S_1$ state although this state has
very large r.m.s. radius, $R\cong 2.5$ fm, even in closed-channel
approximation. Its mass is $M(6\,{}^3S_1)=4815\pm 15$~MeV and
$\Gamma_{ee}(6\,{}^3S_1)=0.20~{\rm keV}$ for $\theta_5=34^\circ$.
The existence of so high a resonance would be important for the
theory.

\section{Conclusions}
\label{sect.8}

We have studied the di-electron widths of higher $n\,{}^3S_1$ and
$n\,{}^3D_1$ radial excitations in charmonium and shown that

\begin{enumerate}
\item
The almost equal values of $\Gamma_{ee}(4040)$ and
$\Gamma_{ee}(4160)$, as well as the small value of
$\Gamma_{ee}(4415)$, can be explained, if large S-D mixing between
$(n+1)\,{}^3S_1$ and $n\,{}^3D_1$ states takes place.
\item
For $\psi(4040)$ and $\psi(4160)$ precise agreement with
experiment is obtained taking the mixing angle
$\theta_2=34.8^\circ$.
\item
For $\psi(4415)$ the calculated di-electron width coincides with
the central value of the experimental width for a mixing angle
$\theta_3=34^\circ$.
\item
In all cases the QCD radiative corrections appear to be important
and the same strong coupling $\alpha_s(M_V)=0.217$ is taken,
giving $\sim 30\%$ effect.
\item
In the single-channel approximation used here DC mass shifts (due
to strong coupling to a nearby threshold) cannot be calculated.
Therefore it remains unclear which physical resonance,
$\psi(4415)$ or the Belle resonance $Y(4360)$, corresponds to the
$3\,{}^3D_1 (4\,{}^3S_1)$ state. For both states we predict close
values of their di-electron widths: $\Gamma_{ee}(\psi(4415))=0.57$
keV and $\Gamma_{ee}(Y(4360)) =0.63(7)$ keV (they coincide within
the experimental error).
\item
Assuming that the $5\,{}^3S_1$ and $4\,{}^3D_1$ states have also
large $S-D$ mixing, with $\theta_4=(32\pm 2)^\circ$), we obtain:
$\Gamma_{ee}(\tilde \psi(4660))=0.35(4)$ keV, $\Gamma_{ee}(\tilde
\psi(4690))=0.40(5)$ keV.
\item
One cannot exclude that a $6\,{}^3S_1$ state also exists, for
which we predict the mass $M(6S)=4815(20)$ MeV and di-electron
width $\Gamma_{ee}=0.20$ keV.
\end{enumerate}

\section{Acknowledgements}
The work is supported by the Grant NSh-4961.2008.2. One of the
authors (I.V.D.) is also supported by the grants of the "Dynasty"
Foundation and the "Russian Science Support Foundation".

\end{document}